\documentclass[a4paper,12pt]{article}
\pdfoutput=1 
\usepackage{jheppub} 
\usepackage{color}
\usepackage{subcaption}

\usepackage{amsmath}
\usepackage[T1]{fontenc} 
\usepackage{mathrsfs}
\usepackage{wrapfig}

\newcommand{\be}{\begin{equation}}
\newcommand{\ee}{\end{equation}}
\newcommand{\beal}{\begin{aligned}}
\newcommand{\eeal}{\end{aligned}}
\newcommand\bea {\begin{eqnarray}}
\newcommand\eea {\end{eqnarray}}
\newcommand{\bec}{\begin{cases}}
\newcommand{\eec}{\end{cases}}
\newcommand{\bei}{\begin{itemize}}
\newcommand{\eei}{\end{itemize}}
\newcommand{\bee}{\begin{enumerate}}
\newcommand{\eee}{\end{enumerate}}

\newcommand{\mum}{\mu_-}
\newcommand{\mup}{\mu_+}
\newcommand{\dmu}{\delta\mu}
\newcommand{\lamm}{\lambda_-}
\newcommand{\lamp}{\lambda_+}

\title{Conical Holographic Heat Engines}

\author[a]{Wasif Ahmed,}
\author[a]{Hong Zhe Chen,}
\author[a]{Elliott Gesteau,}
\author[a,b]{Ruth Gregory,}
\author[b]{Andrew Scoins}

\affiliation[a]{Perimeter Institute, 31 Caroline Street North, Waterloo, 
ON, N2L 2Y5, Canada}
\affiliation[b]{Centre for Particle Theory, Department of Mathematical Sciences
and Department of Physics, Durham University, South Road,
Durham, DH1 3LE, UK}

\emailAdd{wahmed1@perimeterinstitute.ca}
\emailAdd{hchen2@perimeterinstitute.ca}
\emailAdd{egesteau@perimeterinstitute.ca}
\emailAdd{r.a.w.gregory@durham.ac.uk}
\emailAdd{andrew.d.scoins@durham.ac.uk}

\keywords{AdS/CFT, black hole thermodynamics, black hole chemistry, 
holographic heat engines}

\abstract{
We demonstrate that adding a conical deficit to a black hole holographic
heat engine increases its efficiency; in contrast, allowing a black hole to
accelerate \emph{decreases} efficiency if the same average conical deficit
is maintained. Adding other charges to the black hole does not change this 
qualitative effect. We also present a simple formula to calculate the efficiency 
of elliptical cycles for any $C_V\neq 0$ black hole, which allows a more 
efficient numerical algorithm for computation.
}
\preprint{DCPT-19/17}

\begin{document}
\maketitle
\flushbottom


\section{Introduction}

As a consequence of the holographic principle, asymptotically anti-de Sitter 
(AdS) black hole solutions have grown into fruitful playgrounds in which to 
explore aspects of both quantum gravity and strongly coupled gauge field theories
(see e.g.\ \cite{Maldacena:1997re,Gubser:1998bc,Witten:1998qj,Chamblin:1999tk,
Freedman:1999gp,Caldarelli:1999xj,Myers:2010xs}).
Attempts to formulate a consistent thermodynamic description of such objects,
building on the seminal work of Hawking and Page \cite{Hawking:1982dh},
have not only led to relations suggestive of dualities between gravitational 
thermodynamic processes and renormalisation group flows in finite temperature 
quantum field theories \cite{Casini:2011kv,Johnson:2018amj}, they have 
spawned an entirely new field of investigation, known as \emph{black hole chemistry}. 
(For a review this subject, the authors recommend \cite{Kubiznak:2016qmn}).

In the \emph{extended thermodynamics} formalism, a dynamical negative 
cosmological ``constant'' $\Lambda$ is promoted to a thermodynamic 
variable of the black hole \cite{Henneaux:1984ji,Teitelboim:1985dp}, 
interpreted as a positive pressure $P=-\Lambda/8\pi$.
Alongside it's conjugate potential, the \emph{thermodynamic volume} $V$, 
and the usual identifications of horizon area with entropy $S$ and surface 
gravity with temperature $T$, the first law of black hole mechanics takes a
 form reminiscent of the first law of thermodynamics for a more traditional system:
\begin{equation}
dM=TdS+VdP+\ldots .
\end{equation}
The presence of the $VdP$ term (as opposed to $-PdV$) indicates that the 
black hole `mass' $M$, while historically being 
associated with internal energy, has a more correct interpretation of 
\emph{enthalpy} \cite{Kastor:2009wy}, see also \cite{Dolan:2010ha,
Dolan:2011xt,Kubiznak:2012wp,Dolan:2012jh}.

Such a setup has led to the identification of a wide variety of thermodynamic 
phenomena, including entropy inequalities \cite{Cvetic:2010jb,Dolan:2013ft},
Van der Waals-like behaviour 
\cite{Kubiznak:2012wp,Gregory:2019dtq}, 
triple points \cite{Altamirano:2013uqa}, reentrant phase transitions 
\cite{Gunasekaran:2012dq}, and analogous behaviour to superfluidity 
transitions present in condensed matter systems \cite{Hennigar:2016xwd}.
Further, suggestions have been made that pressure variations in the bulk 
might correspond to a variation in a ``chemical potential'' associated to the 
number of colours in the dual CFT \cite{Kastor:2014dra,Dolan:2014cja} 
or to its volume when the number of colours is held fixed \cite{Karch:2015rpa}.

A natural question to pose in light of these developments is whether it is 
possible to construct thermodynamic cycles using these extended 
thermodynamics that one can traverse to extract mechanical work. In a
series of papers, Johnson and collaborators have fleshed out this proposal 
\cite{Johnson:2014yja,Johnson:2016pfa,Chakraborty:2016ssb,
Chakraborty:2017weq,Rosso:2018acz,Johnson:2019olt},
exploring not only the concept of a holographic heat engine, but deriving
their properties, exact expressions and bounds on efficiencies
\cite{Johnson:2014yja,Johnson:2016pfa,Rosso:2018acz}. 
Since the choice of black hole solution provides the 
working substance for such an engine, it is sensible to compare how the 
choice of solution impacts the efficiency of a given cycle.

Na\"{i}vely, one might consider a simple rectangular cycle in the $V$--$P$ 
plane (consisting of two isobars and two isochores) or a Carnot cycle 
(consisting of two isobars and two adiabats) as appropriate for this comparison. 
Indeed, the efficiency of a rectangular cycle for static black holes is readily obtained 
in exact analytic form by Johnson \cite{Johnson:2016pfa},
that was then extended to a general rotating black hole in the canonical 
ensemble by Hennigar et al.\ \cite{Hennigar:2017apu}, who presented 
the efficiency in a simple geometric form.
However, it was suggested by Chakraborty and Johnson that these cycles 
may favour certain black hole solutions \cite{Chakraborty:2016ssb,
Chakraborty:2017weq} 
due to the particular form of the associated equation of state. In order to 
determine which working substances produce the most efficient engines, 
they proposed that a fair comparison can be performed by calculating the 
efficiencies of elliptical cycles instead, as these are in some sense 
``equally unfavourable'' for all black hole solutions.

In this paper, we re-examine this \emph{benchmarking scheme} as applied 
to black holes with conical deficits. Efforts to acquire exact expressions for 
the efficiency of the required elliptical cycles have hitherto been limited to the class of 
solutions with vanishing specific heat at constant volume $C_V=0$ 
\cite{Rosso:2018acz}, for which there can be no rotational charge 
\cite{Hennigar:2017apu} and the sets of isochores and 
adiabats are identical. We show that it is possible to write down an exact 
expression for the efficiency of an elliptical cycle in the $V$--$P$ plane, 
in the canonical ensemble, for a general $C_V\neq 0$ black hole. The 
result takes a pleasant geometric form and provides an efficient algorithm 
with which to calculate efficiency.

Using this result and numerical investigation, we give a discussion of the 
effect of various thermodynamic charges for a class of asymptotically-AdS 
solutions to four dimensional Einstein gravity describing an accelerating, 
rotating, electrically charged black hole. We also consider the impact of 
shifting the position of the benchmarking cycle in the $P$--$V$ plane.
Accelerating black holes in the benchmarking framework have been briefly 
investigated by Zhang et al.\ \cite{Zhang:2018vqs,Zhang:2018hms}, 
although in their study the 
thermodynamics have been unjustly constrained by a choice to remove 
the tension from one of the polar axes. We find that it is the average of the 
``north'' and ``south'' polar tensions that comes to dominate cycle efficiency 
over the acceleration induced by their differential.
The overzealous constraining of the thermodynamic charges has meant 
that this fact has, thus far, gone unnoticed.

The paper is organised as follows: in section \ref{Cmetreview} we review the 
thermodynamics of the metric representing a charged, rotating, slowly-accelerating, 
(asymptotically AdS) black hole. In doing so, we derive a novel bound on the 
metric parameters. In section \ref{Efficiency} we present an exact expression 
for the efficiency of an elliptical benchmarking cycle showing how this provides 
an efficient algorithm to calculate efficiency to a high degree of precision. 
In section \ref{Benchmark} we discuss the benchmarking scheme as applied 
to these accelerating black holes, evaluating the impact of conical deficits 
and extent to which changes in efficiency may be attributed to acceleration.
Finally, in section \ref{concl} we discuss general features of the impact of position 
and geometry of benchmarking cycles on efficiency and conclude.


\section{Thermodynamics of the Charged, Rotating C-Metric}
\label{Cmetreview}

We begin by reviewing the thermodynamics of the charged, rotating 
C-metric \cite{Anabalon:2018ydc,Anabalon:2018qfv}, see also 
\cite{Appels:2016uha,Appels:2017xoe,Gregory:2017ogk,Astorino:2016xiy,
Astorino:2016ybm,MikeThesis,Dutta:2005iy},
which is given in Boyer-Lindquist 
type coordinates \cite{Kinnersley:1970zw,Plebanski:1976gy,Griffiths:2005qp} as
\be
\beal
ds^2 = \frac{1}{H^2} \Bigg[ -&\frac{f(r)}{\Sigma}\left(
\frac{dt}{\alpha}+a\sin^2(\theta)\frac{d\phi}{K}\right)^2
+\frac{\Sigma}{f(r)}dr^2   \\
+&\frac{r^2\Sigma}{g(\theta)}d\theta^2
+\frac{g(\theta)\sin^2(\theta)}{r^2\Sigma}
\left(\frac{a}{\alpha}dt-(r^2+a^2)\frac{d\theta}{K}\right)^2
\Bigg] \,.
\eeal
\label{genCmetric}
\ee
Where the $U(1)$ field strength
\begin{equation}
F=dB\,, \qquad
B=-\frac{e}{r\Sigma}\left[\frac{dt}{\alpha}-a\sin^2\theta\frac{d\phi}{K}\right] 
+\frac{e r_+ dt}{(a^2+r_+^2)\alpha} \,,
\end{equation}
is suitably chosen such that the gauge potential vanishes at the black hole horizon. 
An explicit factor of $K$ has been included so that the azimuthal coordinate 
$\phi$ has periodicity $2\pi$. $K$ typically tracks the presence of conical
deficits in the spacetime, as explained below, thus has physical content, however
$\alpha$, the rescaling of the time coordinate, is not a free parameter in the
metric, but is dependent on the other parameters (as defined in \eqref{alphadef})
and is introduced to appropriately renormalise the timelike killing vector 
$\partial_t$ as described in \cite{Anabalon:2018ydc,Anabalon:2018qfv}. 
The remaining metric functions on this patch are given by
\be
\beal
f(r) &=(1-A^2r^2)\left[1-\frac{2m}{r}+\frac{a^2+e^2}{r^2}\right]
+\frac{r^2+a^2}{\ell^2}\,,\\
g(\theta) &=1+2mA\cos\theta+ (\Xi-1)\cos^2\theta\,,\\
\Sigma &=1+\frac{a^2}{r^2}\cos^2\theta\,, \qquad
H=1+Ar\cos\theta \,,\\
\Xi &= 1 + e^2 A^2 - \frac{a^2}{\ell^2} (1-A^2 \ell^2)\,.
\eeal
\ee

The presence (or not) of conical deficits is revealed by expanding the
angular part of the metric near each axis. Such deficits are interpreted
as cosmic strings emerging from the black hole \cite{Aryal:1986sz}, as
the conical deficits can be smoothed out by a typical cosmic string core
\cite{Gregory:1995hd,Achucarro:1995nu,Gregory:2013xca,Gregory:2014uca}.
The tension of the string, $\mu$, is related to the deficit $\delta$ 
via $\delta = 8\pi \mu$.  Calculating the 
deficits along the North ($+$) and South ($-$) axes gives:
\be
\mu_\pm = \frac{\delta_\pm}{8\pi} = \frac{1}{4}\left[1-\frac{\Xi\pm 2mA}{K}\right]\,.
\ee
As is common, and without loss of generality, we take a non-negative 
acceleration parameter $A$ so that $\mum\geq\mup$.
Often, $\mu_+$ is set to zero so that the North axis is regular, however we 
do not wish to entangle the physics of deficits with the physics of acceleration, 
so will not restrict ourselves thus, but instead will allow both tensions to vary. 
Following \cite{Gregory:2019dtq}, we express the tensions
in terms of the average and differential quantities $\Delta$ and $C$:
\begin{equation}
\Delta = 1-2(\mum+\mup) = \frac{\Xi}{K}
\,,\qquad
\quad C = \frac{\delta\mu}{\Delta} = \frac{\mum-\mup}{\Delta}
=\frac{mA}{\Xi}
\,,
\end{equation}
in order to present the discussion of thermodynamics, although we
use both $C$ and $\delta \mu$ when discussing the impact of acceleration.
Note that these variables are no longer completely unconstrained;
for example, $\dmu$ must vanish for $\Delta=1$. Specifically,
requiring positive tensions and acceleration then gives a bound 
on the magnitude of acceleration: 
\begin{equation}
C<
\begin{cases}
1/2\,, & \text{for $0<\Delta\leq 1/2$,} \\
\frac{1-\Delta}{2\Delta}\,,     & \text{for $1/2<\Delta\leq1$.}
\end{cases}
\end{equation}

It is worth reiterating the relation between the physics of acceleration and 
that of deficits. A black hole can have a conical deficit without accelerating,
and $\Delta$ encodes this property, however, note that $\Delta=1$ for zero 
deficit and acceleration, then \emph{drops} as the conical deficit average 
\emph{increases}. $C$ increases as acceleration increases, saturating
at $C=1/2$ for a critical black hole, defined as having a deficit of $2\pi$ along
the South axis. Since it is not possible for the deficit to increase further,
$C$ remains at $1/2$ independent of the North pole deficit, but the 
\emph{acceleration} of the black hole \emph{drops} as $\mu_+$ increases,
returning to zero as $\mu_+\to\mu_-$ and $\Delta\to0$. Therefore, while $C$
is a convenient parameter to express the extensive thermodynamical
variables \cite{Gregory:2019dtq}, $\delta\mu$ is more representative of 
acceleration and we will frequently use it for displaying results.

At this point, it is worth commenting on the parametric restrictions in the metric.
Positivity of the conformal factor $H = 1+Ar\cos\theta$, constrains 
$Ar\cos\theta<1$ and sets the location of the conformal boundary 
$r_\text{bd.}=-1/A\cos\theta$.
Our main assumption is that the black hole is {\it slowly accelerating},
(see \cite{Podolsky:2002nk} for a full discussion) i.e.\ its
time coordinate is proportional to the asymptotic time for an observer near the 
boundary. For the black hole to also be isolated (i.e.\ the only event horizon being
that of the black hole) we require no zeros of $g_{tt}$, or $f$, on the boundary.
Finally, for $\theta=0,\pi$ to represent the poles, we require 
$g(\theta)>0$ on $[0,\pi]$. All these requirements lead to a set of 
intersecting constraints on the parameters.

First, $g\geq0$ gives a bound on the possible values of the dimensionless mass: 
\be
mA< \begin{cases}
\Xi/2,          & \text{for $\Xi\in(0,2]$,} \\
\sqrt{\Xi-1},   & \text{for $\Xi>2$.}
\end{cases}
\label{mAconditions1}
\ee
However, the fact that the black hole horizon does not intersect the boundary 
requires $Ar_+<1$, hence the Kerr-Newman potential multiplying $(1-A^2r_+^2)$
in $f(r_+)=0$ must be negative. This in turn requires 
\be
m^2 > a^2 + e^2 \qquad\Rightarrow\qquad
m^2 A^2 > \Xi -1 + \frac{a^2}{\ell^2} > \Xi-1
\,.
\ee
Thus, by comparison with \eqref{mAconditions1}, we see that $\Xi>2$ is not allowed.
Hence
\be
mA \leq \frac{\Xi}{2} <1.
\ee

The condition for slow acceleration, or $f>0$ on the boundary, then
becomes an algebraic constraint on the parameters, that must be satisfied
in conjunction with the existence of a black hole horizon. This is a rather
involved set of constraints which are most easily solved numerically. 
We refer the reader to \cite{MikeThesis} for a fuller discussion.

Consistent thermodynamic parameters for this class of solutions have 
been identified in \cite{Anabalon:2018qfv} by promoting $\mu_\pm$ to 
thermodynamic charges, taken together with their conjugate 
\textit{thermodynamic lengths} $\lambda_\pm$. We restate them here for clarity:
\be
\beal
M&= \frac{m(\Xi+a^2/\ell^2)(1-A^2 \ell^2\Xi)}{K\Xi\alpha(1+a^2A^2)}\,,\\
T&= \frac{f'_+ r_+^2}{4\pi\alpha(r_+^2+a^2)}\,, \qquad
S=\frac{\pi(r_+^2+a^2)}{K(1-A^2r_+^2)}\,,\\
Q&= \frac{e}{K}\,,\quad \Phi=\Phi_t=\frac{er_+}{(r_+^2+a^2)\alpha}\,,\\
J& =\frac{ma}{K^2}\,,  \quad \Omega=  \Omega_H-\Omega_\infty\,
=\left ( \frac{Ka}{\alpha(r_+^2+a^2)}\right ) -
\left ( -\frac{aK(1-A^2\ell^2\Xi)}{\ell^2\Xi \alpha(1+a^2A^2)}\right) \,,\\
P &= \frac{3}{8\pi \ell^2} \,, \quad
V = \frac{4\pi}{3K\alpha} \left [ \frac{r_+(r_+^2 + a^2)}{(1-A^2 r_+^2)^{2}}
+ \frac{m[a^2(1-A^2\ell^2 \Xi) + A^2 \ell^4 \Xi (\Xi+a^2/\ell^2)]}
{(1+a^2 A^2) \Xi} \right]\,,\\
\lambda_\pm &= \frac{-r_+}{\alpha(1\pm Ar_+)} +\frac{m}{\alpha}
\frac{[\Xi + a^2/\ell^2 +   \frac{a^2}{\ell^2} (1-A^2\ell^2 \Xi)]}{(1+a^2 A^2)\Xi^2}
\pm \frac{A \ell^2 (\Xi +  a^2/\ell^2 )}{\alpha(1+a^2A^2)}\,,
\eeal
\label{thermo_metricfunctions}
\ee
with the correct normalisation of the timelike killing vector given by
\begin{equation}
\alpha=\frac{\sqrt{(\Xi+a^2/l^2)(1-A^2l^2\Xi)}}{1+a^2A^2}\,.
\label{alphadef}
\end{equation}
The parameters (\ref{thermo_metricfunctions}) were shown to satisfy both 
an extended first law of thermodynamics
\begin{equation}
dM  =  TdS+VdP    +\Omega dJ+\Phi dQ    +\lamp d\mup +\lamm d\mum\,,
\end{equation}
and Smarr relation \cite{Smarr:1972kt}
\begin{equation}
M    =    2(TS-PV+\Omega J)    +\Phi Q\,.
\end{equation}
Interesting thermodynamic behaviour -- including zeroth, first, and second 
order phase transitions and the first example of a reentrant black hole 
phase transion when $P$ is varied -- of these solutions has been discussed 
in the literature~\cite{Abbasvandi:2018vsh,Abbasvandi:2019vfz}. 
It is also possible to rewrite the expressions (\ref{thermo_metricfunctions}) 
in terms of the thermodynamic charges~\cite{Gregory:2019dtq}:
\be
\beal
V &=
\frac{2S^2}{3\pi M} \left[ \left(1+\frac{\pi Q^2}{\Delta S}+\frac{8PS}{3\Delta}\right)
+ 2\left(\frac{\pi J}{\Delta S}\right)^2 +2\left(\frac{3\Delta C}{8PS}\right)^2
\right] \,,\\
T &= \frac{\Delta}{8\pi M}
\Bigg[\left(1+\frac{\pi Q^2}{\Delta S}+\frac{8PS}{3\Delta}\right)
\left(1-\frac{\pi Q^2}{\Delta S}+\frac{8PS}{\Delta}\right )
-4\left(\frac{\pi J}{\Delta S}\right)^2	-{4C^2}
\Bigg]
\,,\\
\Omega &= \frac{\pi J}{S M \Delta} \left(1+\frac{8PS}{3\Delta}\right) \,,\\
\Phi &= \frac{Q}{2M}\left(1+\frac{\pi Q^2}{S\Delta}+\frac{8PS}{3\Delta}\right)\,,\\
\lambda_\pm    &=
\frac{-S}{\pi M}
\Bigg[ \! \!\left ( \!\frac{4PS}{3\Delta} + \frac{\pi Q^2}{2\Delta S} \!\right)^2
\!\!+ \frac{\pi^2 J^2}{\Delta^2 S^2} \left (\!1+ \frac{16PS}{3\Delta}\!\right)
\!-\left ( 1 \mp {2C}\right)^2 \pm 4\left(\frac{3\Delta C}{8PS}\right)\!
\Bigg]
\eeal
\label{TDvars}
\ee
giving a generalisation of the Christodoulou-Ruffini formula
\cite{Christodoulou:1972kt, Caldarelli:1999xj} for enthalpy:
\be
M^2 =
\frac{\Delta S}{4\pi}
\left[ \left(1+\frac{\pi Q^2}{\Delta S} +\frac{8PS}{3\Delta}\right)^2
+4\left(1+\frac{8PS}{3\Delta}\right)
\left\{\left(\frac{\pi J}{\Delta S}\right)^2 - \frac{3\Delta C^2}{8PS}\right\}
\right]
\,.
\label{ChRu}
\ee
Having expressions purely in terms of extensive quantities clarifies the
chemical interpretation of black holes, and enables a more
straightforward analysis of the properties of accelerating heat engines.

These expressions make it clear that the reverse isoperimetric inequality
\cite{Cvetic:2010jb} is satisfied for this class of solutions 
\cite{Gregory:2019dtq}.


\section{An Exact Efficiency Formula for Circular Cycles}
\label{Efficiency}

As discussed in the introduction, previous authors have calculated the 
efficiency of elliptical benchmarking cycles for black hole solutions with $C_V=0$.
However, it is possible to make a more general statement, extending this 
result to a broader class of solutions.

Recall that an engine consists of a cycle that has a ``cool'' component, where 
work is extracted, and a ``hot'' component, where the engine is refuelled. The 
efficiency is simply the ratio of overall heat extracted to the heat put in,
where the heat flow is given by an integral
\be
Q = \int TdS 
\ee
over each component of the cycle. 
The transition between the hot and cold parts of the cycle occurs when
$\delta S=0$. Thus, on any cycle in the $(V,P)$ plane, the turning points of
$S$ must be determined. The trickiness of the problem now becomes
apparent, as the thermodynamic variables are most readily given in terms of the 
charges $S,P,\ldots$, whereas we require $M(P,V)$ so that we can determine the
stationary points of $S$.

\begin{figure}
\centering
\includegraphics[width=0.6\textwidth]{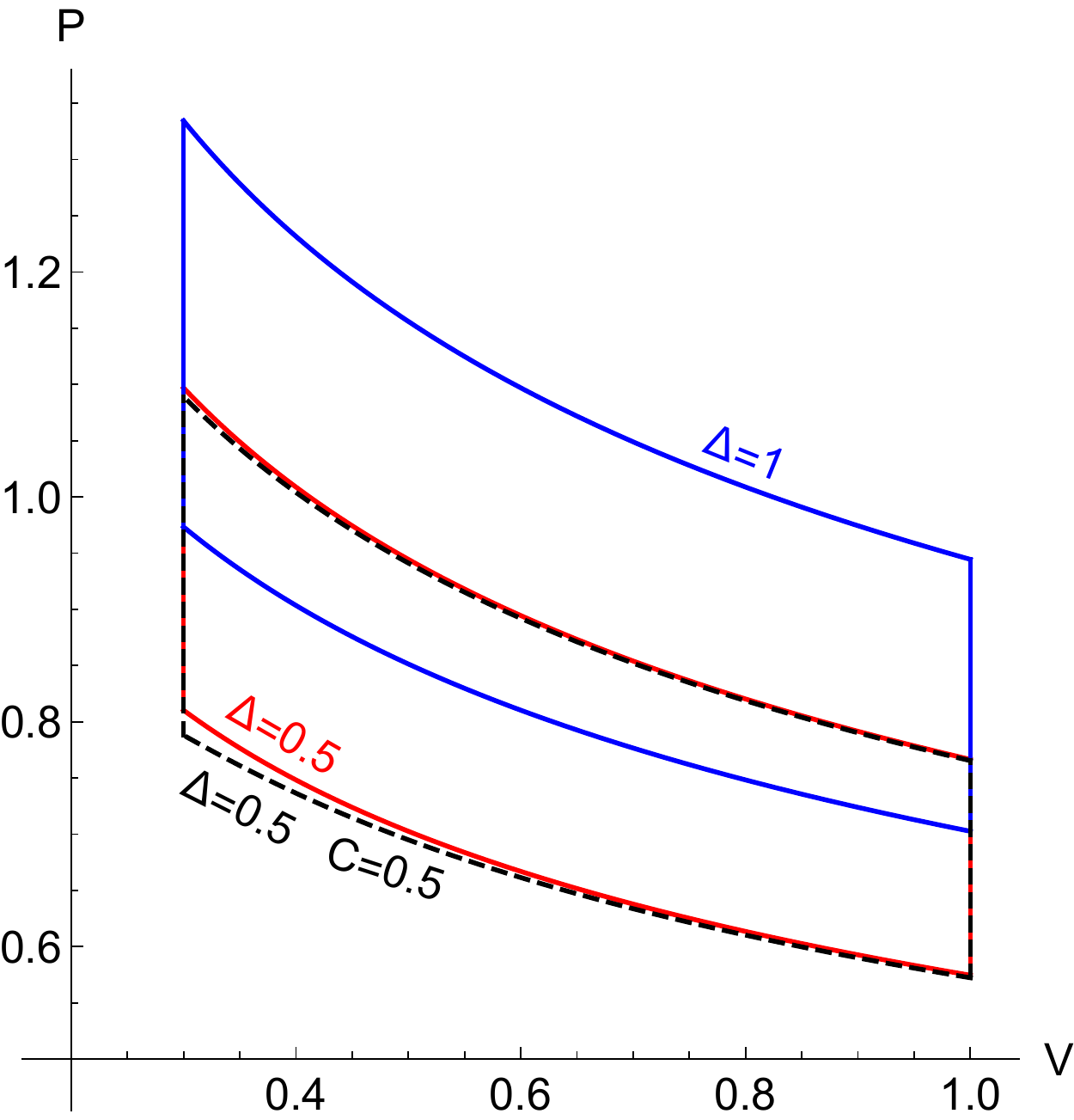}
\caption{An illustration of the impact of conical deficits and acceleration
on the Carnot cycle of an uncharged nonrotating black hole.The cycles 
have upper (lower) temperature $T=1.3$ ($1.0$) 
and maximum (minimum) volume $V=1.0$ ($0.3$). First an average
deficit $\Delta=0.5$ is added (solid red), and can be seen to have a strong
effect on the geometry of the cycle, however, when the maximum
possible thermodynamic acceleration, $\delta\mu=0.25$, $C=0.5$, is
added (still having the same \emph{average} deficit $\Delta=0.5$)
the cycle changes very little.
}
\label{fig:scaling}
\end{figure}

Two common simple cycles used are the Carnot,
and a $(V,P)$--rectangular cycle. These have clear turning points for $S$
at the corners. However, while the rectangular cycle is easy to use
for black holes with conical deficits, it turns out that the Carnot cycle is
not. The Carnot cycle consists of adiabats and isotherms. The engine is
first heated up, by increasing the pressure at fixed volume, then allowed to
expand at constant temperature. The cycle then completes by cooling to 
the original temperature by dropping the pressure, and contracting to the 
original $P$ and $V$ at constant $T$. The geometry of the Carnot cycle 
therefore depends crucially on the isotherms of the system. Looking at
\eqref{ChRu} we see a scaling symmetry,
\be
\hat {S} =  \Delta S\,,\qquad
\hat{P} = P/ \Delta^2\,,
\label{spscale}
\ee
so that
\be
V =\frac{\hat{V} ( \hat{S}, \hat{P} )}{ \Delta^2 } \,,\qquad
T = \Delta \hat{T}( \hat{S}, \hat{P} )\,,
\label{tvscale}
\ee
where $\hat{T}$, $\hat{V}$ are the expressions for the undeficited black
hole. 
Thus, irrespective of any impact of acceleration, turning on a conical
deficit gives a transformation on the $(V,P)$ plane, and
not only distorts, but rescales, the isotherms. For a Carnot engine 
cycling between the same two temperatures and entropies, the addition of a 
deficit ``drops the pressure'' of the cycle, lowering the maximum pressure 
obtained. By looking at the expressions \eqref{TDvars} for $T$ and $V$, one 
can deduce that both quantities are only significantly altered by the addition 
of the terms containing $C$ when both $T$ and $V$ are small. These
effects are depicted in figure \ref{fig:scaling}, showing Carnot engines for a
``vanilla'' (uncharged nonrotating) black hole cycling between the same two 
temperatures and entropies. First an average
deficit $\Delta=0.5$ is added (solid red), and can be seen to have a strong
effect on the geometry of the cycle, however, when the maximum
possible thermodynamic acceleration, $\delta\mu=0.25$ or $C=0.5$, is
added (still having the same \emph{average} deficit $\Delta=0.5$)
the cycle changes very little. As is the case generally, the presence of the 
deficit dominates the phase-space position and shape of the cycle, with the 
deformation due to acceleration only becoming non-negligible at small $V$ and $P$. 

Since introducing a deficit distorts and rescales the isotherms needed to construct 
Carnot cycles, it is necessary to instead use a geometrically fixed benchmarking cycle.
Rectangular cycles are relatively straightforward. Indeed, an exact formula 
for their efficiency for any black hole was given in 
\cite{Hennigar:2017apu}:
\be
\eta = \frac{\Delta V\Delta P}{\Delta M_T +\Delta U_L}
\,,
\ee
where $\Delta M_T$ is the difference in enthalpy between the top-right and 
top-left corners of the cycle in the $(V,P)$-plane, and $\Delta U_L$ is the 
difference in internal energy $U=M-PV$ between the top-left and bottom-left corners.
For the vanilla C-metric, this evaluates to
\be
\eta = \frac{2A}{A+2P_0\Delta V + \left(\frac{3\Delta^2}{4\pi}\right)^\frac13
\left(    V_R^\frac13-V_L^\frac13   \right)}
+\mathcal{O}(C^2)
\,,
\ee
where $A=\Delta V\Delta P$ is the area of the cycle; $V_R$ and $V_L$ are the maximum 
and minimum volumes attained respectively; and $P_0$ is the pressure at the cycle's 
centroid. The correction of order $C^2$ is calculated by resorting to the expansion 
\eqref{notsosimpleS} which we present later in our analysis of circular cycles. 

\begin{figure}
\center{\includegraphics[width=7cm]{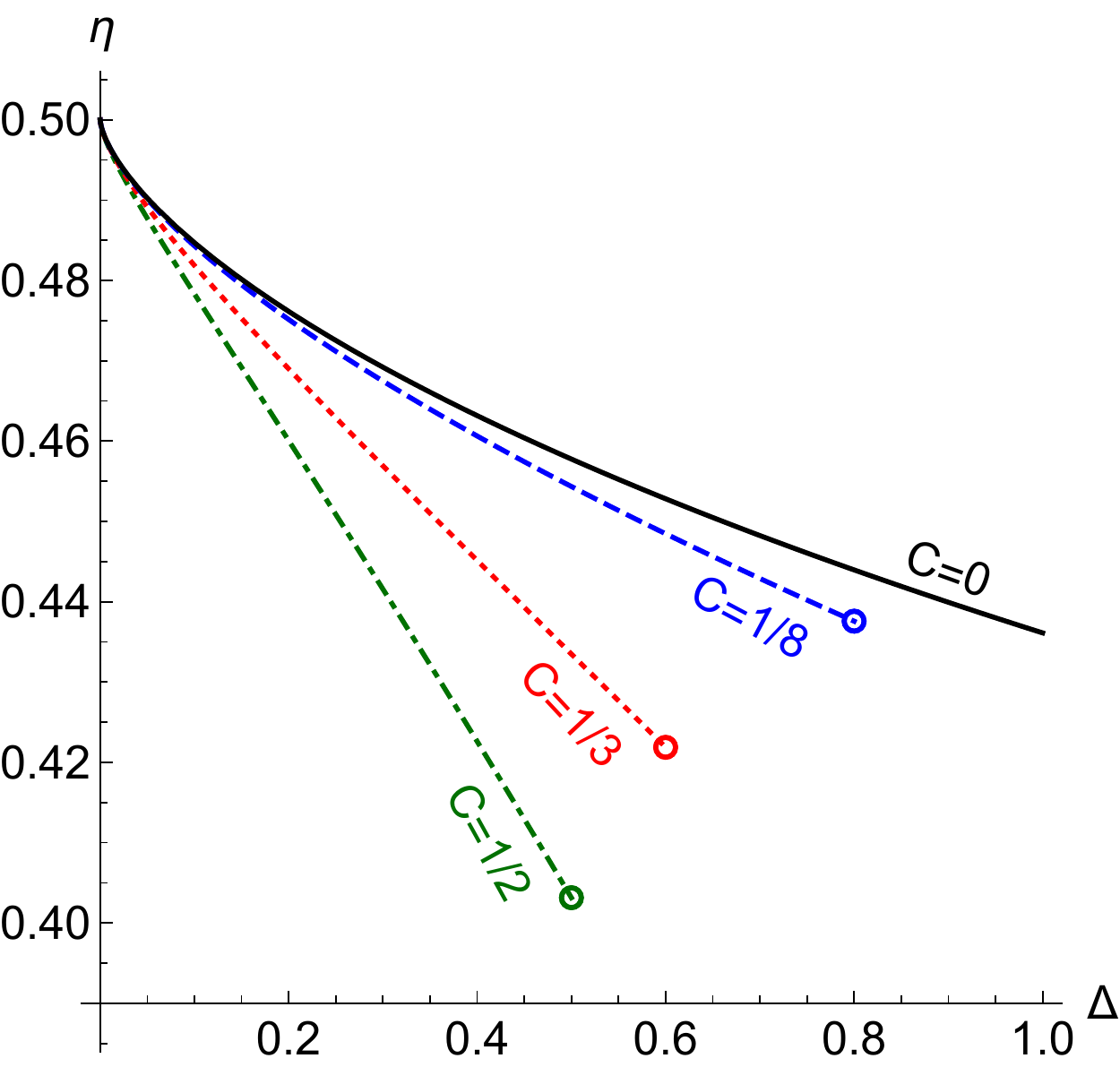}}
\caption{The efficiency of a typical rectangular cycle with maximum 
(minimum) pressure $P=1.0$ ($0.5$) and volume $V=1$ ($0.5$), 
for a $Q=J=0$ solution. The non-accelerating case is shown in solid 
black. Cases with three values of non-zero $C$ are plotted using broken 
curves, up to the maximum legal $\Delta$. Increasing the average deficit 
(dropping $\Delta$) is seen to increase efficiency. Greater values of 
acceleration are seen to give reduced efficiency.}
\label{fig:rectangleEfficiency}
\end{figure}

Adding a deficit (i.e.\ \emph{lowering} $\Delta$) acts to increase the
efficiency of rectangular cycles. When calculated, the  correction from 
acceleration decreases the efficiency again, albeit by a lesser amount. 
These effects are demonstrated in figure \ref{fig:rectangleEfficiency} wherein the 
efficiency of a typical rectangular cycle is plotted against $\Delta$ for various 
values of $C$. Here we see how increasing the deficit (corresponding
slightly counterintuitively by moving to the left in the plot) increases
efficiency. Adding acceleration via $C$ on the other hand lowers
efficiency, as can be seen from the coloured dashed/dotted lines
lying below the black $\delta\mu=0$ curve.

As discussed in the introduction, previous authors 
\cite{Rosso:2018acz,Chakraborty:2016ssb,Chakraborty:2017weq} 
have also calculated the 
efficiency of circular benchmarking cycles for black hole solutions with $C_V=0$.
However, we can use the first law to give a more general
geometric expression for the efficiency for any black hole heat engine, even those 
with non-vanishing specific heat.
Consider for convenience a circular benchmarking cycle defined parametrically by
\be\beal
V(\theta)   &=  V_0 + R\cos{\theta}     \,,\\
P(\theta)   &=  P_0 - R\sin{\theta}     \,.
\eeal
\label{pvcircle}
\ee
Here, $(V_0,P_0)$ indicates the centre of the cycle and $R$ its radius\footnote{
An elliptical cycle may be put in this form by rescaling the units of the dimensionful 
quantities $P$ and $V$. This amounts to giving different values of $R$ for each.
}.
Calculating the work done in traversing the circle is in
principle straightforward as, in the canonical ensemble, the heat flow
is given simply by the first law:
\begin{equation}\begin{aligned}
\delta Q    &=  TdS   =   dM-VdP  \,,\\
\implies Q  &=  \Delta M - \int VdP    \,.
\end{aligned}
\label{workdone}
\end{equation}

\begin{figure}
\center{\includegraphics[width=7cm]{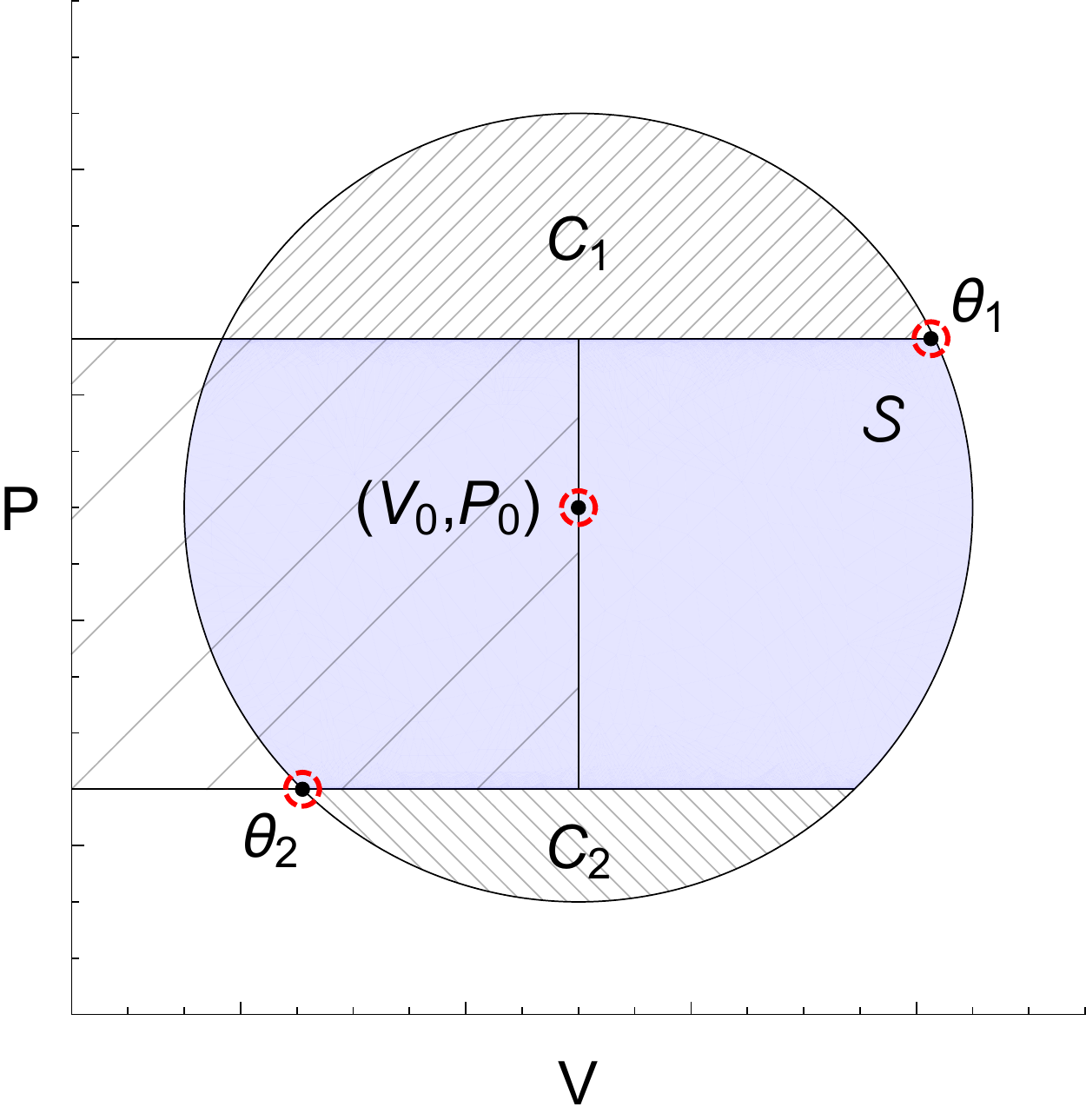}}
\caption{A general benchmarking cycle, partitioned into subregions.}
\label{fig:integralDiagram}
\end{figure}

In general the presence of charges means that isochores are no longer adiabats, 
and the turning points of $S$ along the circle are displaced from the symmetric 
position (see figure \ref{fig:integralDiagram}), however the first law integral 
\eqref{workdone} is still applicable and we may write the integrals for $Q_C$ 
and $Q_H$ as simple combinations of mass differentials and areas of regions in 
the circle. Assuming the turning points $\theta_{1,2}$ divide the circle into 
two segments, with areas $C_1$ and $C_2$, above and below the chords parallel 
to the $V$--axis defined by $\theta_1$ and $\theta_2$ respectively,
the strip of the circle remaining has area ${\cal S} = \pi R^2- C_1-C_2$.
These regions are shown in figure \ref{fig:integralDiagram}.
Inspection of the area
under the $V$--curve to the $P$--axis then gives the expressions
\be
\beal
-Q_C = M_2-M_1 + V_0(P_1-P_2) + \frac12 {\cal S} + C_2 \,, \\
Q_H = M_1-M_2 - V_0(P_1-P_2) + \frac12 {\cal S} + C_1 \,,
\eeal
\label{heatflowsgen}
\ee
allowing one to straightforwardly write down the efficiency:
\be
\eta
=\frac{\pi R^2}{M_1-M_2 -V_0 R(\sin\theta_2-\sin\theta_1)
+ \frac{R^2}{2}[\theta_1-\theta_2 + \sin(\theta_1-\theta_2)
\cos(\theta_1-\theta_2)]}
\,.
\label{exactefficiency}
\ee
The turning points are typically found numerically, and the mass determined by
solving for $S(V,P)$ at $\theta_1, \theta_2$, and inputting into $M$.
Note that this method is extremely efficient numerically, as one can 
discretise the circle very coarsely to get a ballpark range for the $\theta_i$, 
then refine for the precision required. The numerical problem is fairly 
independent of the circle size, and is mostly dependent on the level of
precision desired.

One should note that for cases of vanishing $C_V$, (such as a black hole 
described by the uncharged C-metric), $(\theta_2,\theta_1)$ aproaches $(0,\pi)$ 
and our efficiency formula \eqref{exactefficiency} reduces to the previously 
found expression \cite{Hennigar:2017apu}:
\be
\eta=\frac{\pi R^2}{\pi R^2/2 +\Delta M}
\,.
\label{CVzeroeta}
\ee


\section{Impact of Conical Deficits on Cycle Efficiency}
\label{Benchmark}

Now we would like to explore the impact of conical deficits and acceleration
on the efficiency of holographic heat engines. We consider a circular cycle of the 
type described by \eqref{pvcircle}.

To determine the turning points of $S$, we must invert the expression 
$V(S,P)$ to obtain $S(P,V)$. This is a conceptually straightforward, though
algebraically involved, procedure, complicated however by
the fact that with acceleration $S(V,P)$ is multivalued, leading to a
constraint on parameter space discussed in \S \ref{concl}. 
Fixing $Q$, $J$, $\Delta$, and $\delta\mu$
and substituting the expressions \eqref{pvcircle} into the volume
\be
V(\theta) = \frac{4S^{3/2}}{3\sqrt{\pi\Delta}} \frac{\left[ 1+\frac{\pi Q^2}{\Delta S}
+\frac{8P(\theta) S}{3\Delta}
+ 2\left(\frac{\pi J}{\Delta S}\right)^2 +2\left(\frac{3\Delta C}{8P(\theta)S}\right)^2
\right]}{\left[ \left(1+\frac{\pi Q^2}{\Delta S} +\frac{8P(\theta)S}{3\Delta}\right)^2
+4\left(1+\frac{8P(\theta)S}{3\Delta}\right)
\left\{\left(\frac{\pi J}{\Delta S}\right)^2 - \frac{3\Delta C^2}{8P(\theta)S}\right\}
\right]^{1/2} 
} \,
\ee
leads to a rational (though complicated!) expression for $S(\theta)$.
Once we have the expressions for heat flow in terms of $P$ and $V$,
a further constraint arises from requiring that the black hole indeed does
have a horizon -- i.e.\ that the rotation or charge is below or at the extremal
limit. 

Insight into the behaviour with deficits and acceleration can be gained
by considering the simple case $Q=J=0$. Setting $C=0$ at first, things 
simplify considerably, and 
\be
S = \left ( \pi\Delta\right)^{\frac13} \left ( \frac{3V}{4} \right)^{\frac23} 
\;\;\;\Rightarrow\quad M = PV + \frac{\Delta^{\frac23}}{2}
\left ( \frac{3V}{4} \right)^{\frac13}
\,.
\label{simpleS}
\ee
In this very simple case, $\delta S=0$ at the turning points of the circle
$\delta V=0$, i.e.\ $\theta=0,\pi$, and the integral of $VdP$ around each half
of the cycle simply gives half of the area of the circle, $\pi R^2/2$. Thus, the
efficiency takes the straightforward form
\be
\eta
= \frac{2 \pi R^2}{ \pi R^2 + 4RP_0 +  (\frac{3\Delta^2}4)^{\frac13}
\left ( (V_0+R)^{\frac13} -(V_0-R)^{\frac13} \right)}
\,.
\label{nochargeseta}
\ee
An ideal gas black hole has $M=PV$, and the efficiency of its benchmarking cycle 
has been evaluated \cite{Hennigar:2017apu} to be
\be
\eta_\text{ideal gas}
=
\frac{2\pi R^2}{\pi R^2+4RP_0}
\,.
\ee
Such a solution can (thermodynamically) be regarded as the limit 
of the Schwarzschild-AdS metric as horizon radius grows, 
and acts as an upper bound on efficiency \cite{Chakraborty:2016ssb}.
One can see from \eqref{nochargeseta} that the essential effect 
of $\Delta$ is to damp down the ``non-ideal gas'' part of the 
black hole behaviour; black holes with a conical deficit 
approach the ``large black hole'' limit more rapidly than their 
non-deficited counterparts.
\begin{figure}
\center{\includegraphics[width=7cm]{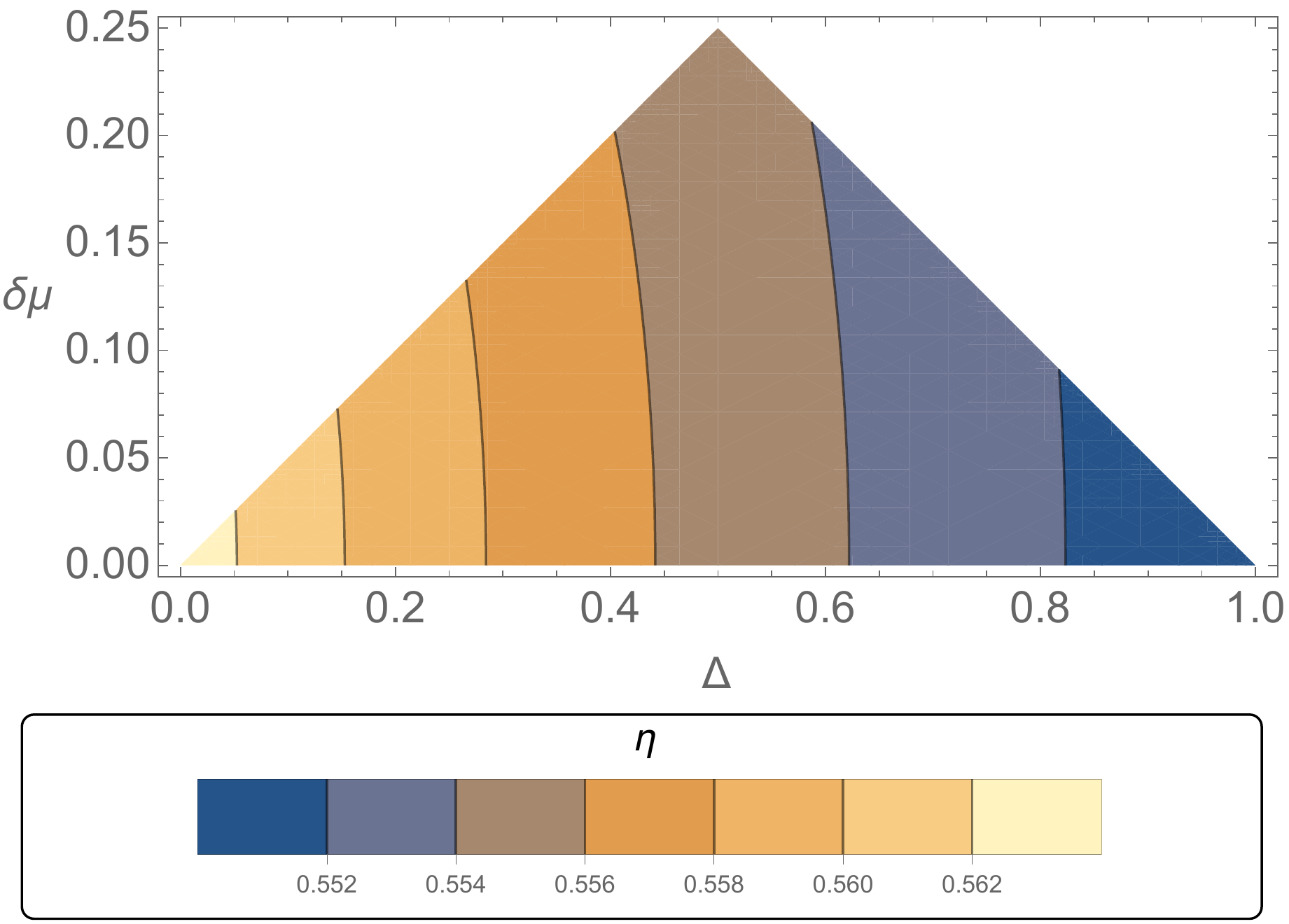}}
\caption{A contour plot to illustrate the effect of deficits and acceleration
on uncharged nonrotating solutions. The efficiency is shown as a function
of $\Delta$ and $\delta\mu$ for a cycle of unit radius centred at $(V_0=2,P_0=2)$.}
\label{fig:nocharges}
\end{figure}

Let us repeat our analysis to assess the effect of acceleration. Including 
non-zero $C$ in \eqref{simpleS} yields
\be
\beal
V &= \frac{4S^{3/2}}{3\sqrt{\pi\Delta}} \frac{1
+\left(\frac{3\Delta}{8PS}\right)^2\frac{2C^2}{(1+8PS/3\Delta)}}
{\left [ 1-\left(\frac{3\Delta}{8PS}\right)\frac{4C^2}{(1+8PS/3\Delta)}\right]^{1/2}}\\
&\approx  \frac{4S^{3/2}}{3\sqrt{\pi\Delta}} \left [ 1
+2\left(\frac{3\Delta C}{8PS}\right)^2 \right]
\eeal
\ee
upon expansion to order $C^2$. Note that since each tension $\mu_\pm$ is bounded
by $1/4$, $\Delta C\leq1/4$. Hence, unless we are dealing with very small
black holes, this should be an excellent approximation. Inverting,
we find
\be
\beal
S &= \left ( \pi\Delta\right)^{\frac13} \left ( \frac{3V}{4} \right)^{\frac23}
\left [ 1- \frac{2C^2}{x_0^2}\right] ^{\frac13}\\
\Rightarrow\quad M &= PV + \frac{\Delta^{\frac23}}{2}
\left ( \frac{3V}{4\pi} \right)^{\frac13}
- \frac{\Delta^2 C^2}{16 \pi P^2 V} \left ( 1
+ 8\pi P\left( \frac{4V^2}{3\pi^2\Delta^2} \right)^{\frac13}
\right)
\,,
\eeal
\label{notsosimpleS}
\ee
where for shorthand we write
\be
x_0 = \frac{8PS}{3\Delta} \Bigg|_{C=0} 
= 2P \left ( \frac{4\pi V^2}{3\Delta^2} \right)^{\frac13}
\,.
\ee
Thus, the difference in mass across the cycle is 
\be
\Delta M = P_0 \Delta V + \frac{\Delta^{\frac23}}{2}
\left ( \frac{3}{4\pi} \right)^{\frac13} (V_1^{\frac13}-V_2^{\frac13})
+ \frac{\Delta^2 C^2}{16 \pi P_0^2} \left ( \frac{V_1-V_2}{V_1V_2}
+ 6^{\frac23} P_0
\frac{4(V_1^{\frac13}-V_2^{\frac13})}{\Delta^{\frac23}V_1^{\frac13}V_2^{\frac13}}
\right)
\,.
\ee
From \eqref{CVzeroeta} the effect of acceleration is to decrease efficiency, 
whereas the effect of a conical deficit is to increase it. This is illustrated 
in figure \ref{fig:nocharges}, and refines the 
findings of \cite{Zhang:2018vqs} in which it was (incorrectly) reported that 
acceleration increases efficiency. Their conclusion followed from a choice of 
$K$ which regularised one of the C-metric's poles. However, this constrains the 
thermodynamic charges and removes the independence of $\Delta$ and $C$. 
Our analysis shows that the situation is in fact more subtle. The dominant 
effect is the existence of the conical deficit, not the acceleration itself.
When the thermodynamic charges are constrained by fixing $K$, these two effects 
are inseparable, giving a misleading picture of the phenomenology. This viewpoint 
is concurrent with the findings of \cite{Gregory:2019dtq} in which it was argued that 
the average deficit is usually the more impactful of the two effects for the 
thermodynamics.

Figure \ref{fig:nocharges} shows the efficiency contours in the $(\Delta,\delta\mu)$
plane for a circular benchmarking cycle of unit radius centred on $P_0=V_0=2$.
The relatively small values of $P_0$ and $V_0$ (in contrast to the benchmarking
cycles of \cite{Hennigar:2017apu,Zhang:2018vqs}) were chosen to maximise the
impact of acceleration. Even so, the contour lines are predominantly vertical, 
indicating that it is the value of $\Delta$, the mean deficit, that dominates the
change in efficiency.

Once charges are added, the level of complexity rapidly rises, not least
because the turning points from hot to cool parts of the cycle now arise at
different $P$. As  $Q$ or $J$ are increased, the points $\theta_i$ at which
$\delta S$ vanishes are ``shifted clockwise'' around the cycle, in a qualitatively
similar arrangement to the one shown in figure \ref{fig:integralDiagram}.
We numerically explored this shifting, identifying $\theta_i$ for the complete
range possible of cycle radii (those which kept pressure and volume positive)
for a range of $P_0$ and $V_0$ from close to zero to values of order $10^3$,
but were unable to identify valid choices of charges or deficits for which 
$\theta_1$ and $\theta_2$ did not fall in the upper--right and lower--left 
quadrants of the cycle respectively. This verifies that our efficiency formula
\eqref{exactefficiency} is valid across this broad sampling of the phase space 
for the metric \eqref{genCmetric}.
\begin{figure}
\centering
\begin{subfigure}[b]{0.45\textwidth}
\includegraphics[width=\textwidth]{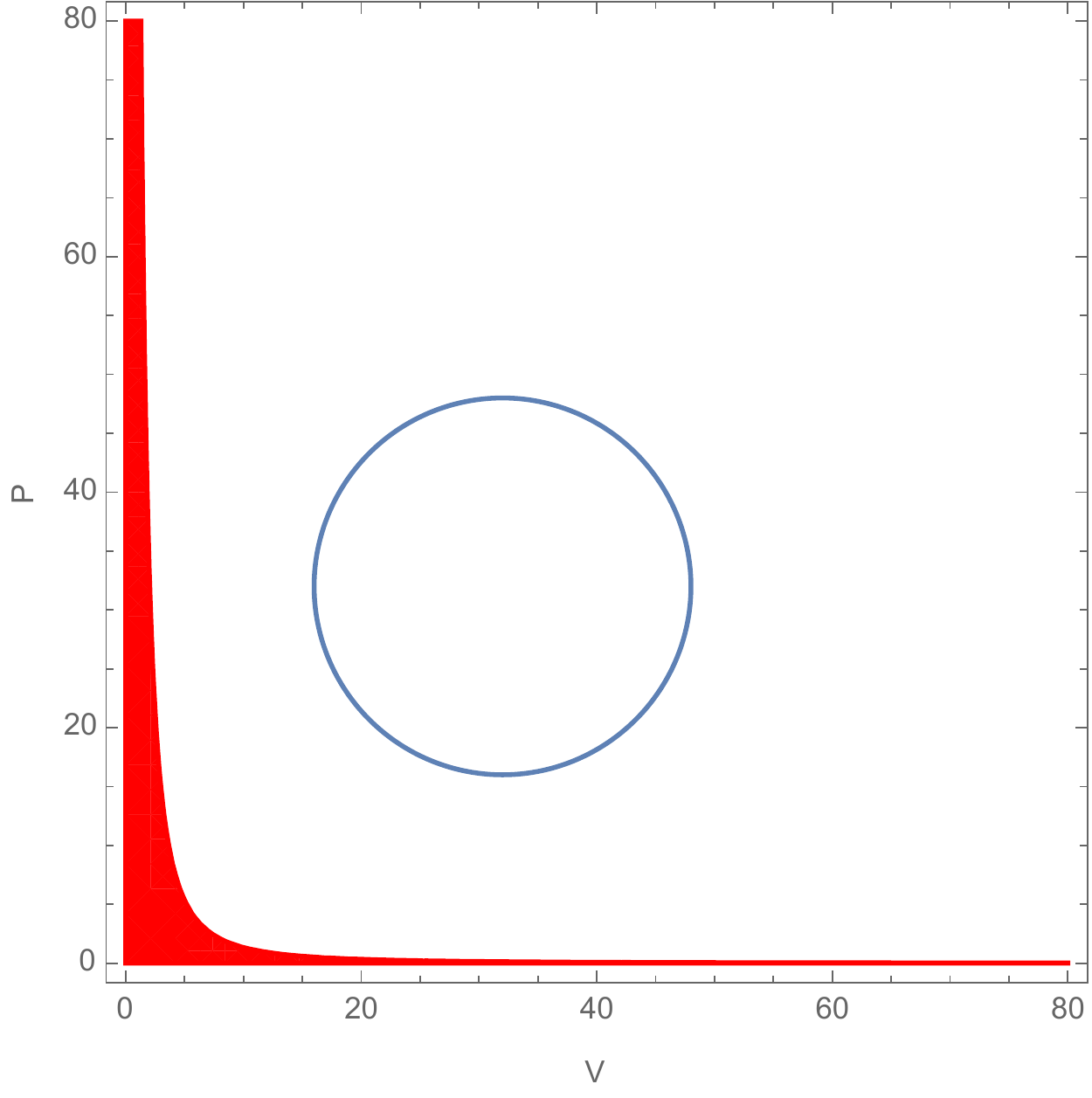}
\caption{\label{fig:PVplaneJ8noD} $J=8$, $Q=0$, $\Delta=1$, $\delta\mu=0$.}
\end{subfigure}
\quad
\begin{subfigure}[b]{0.45\textwidth}
\includegraphics[width=\textwidth]{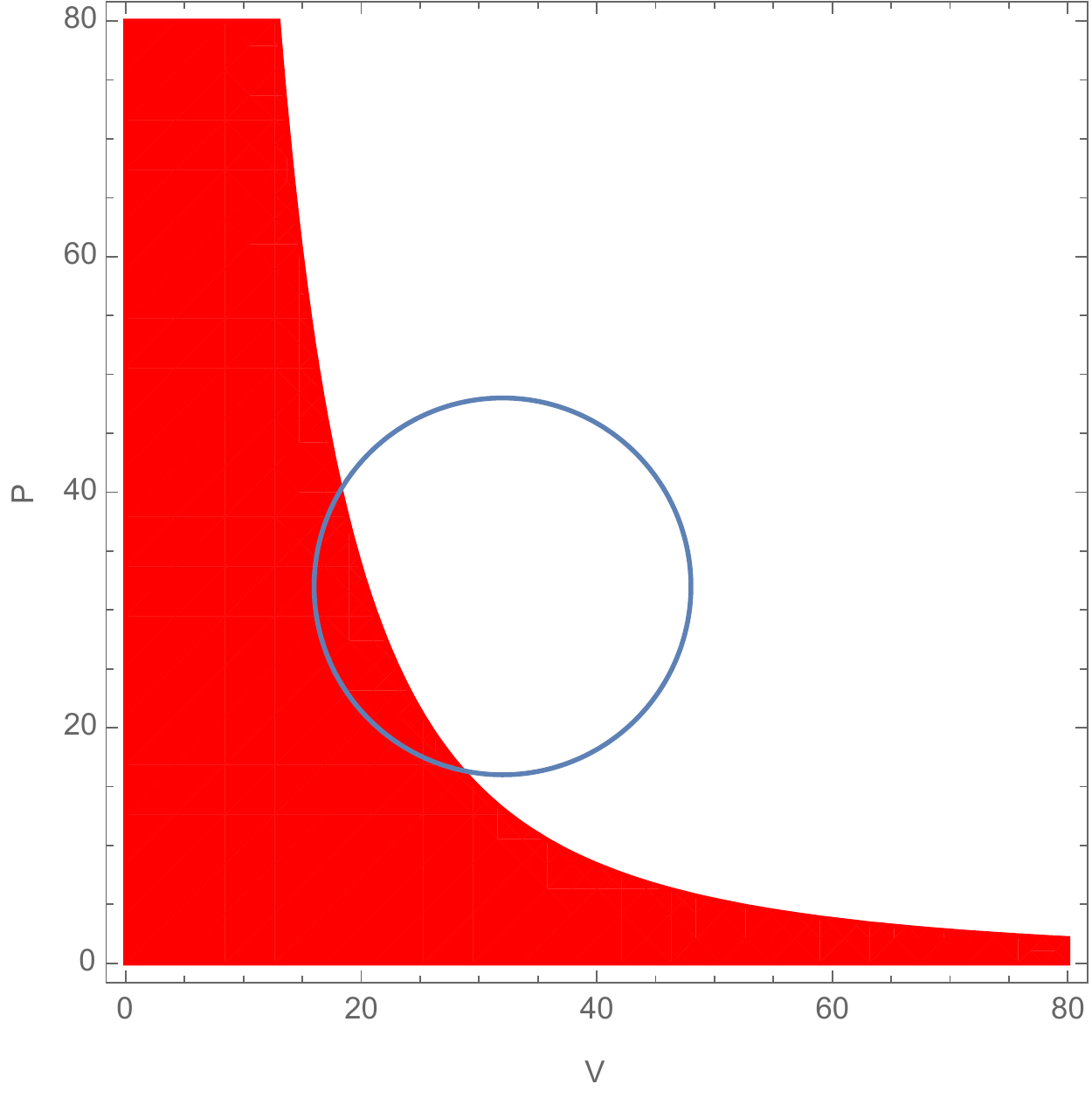}
\caption{\label{fig:PVplaneJ8D} $J=8$, $Q=0$, $\Delta=0.1$, $\delta\mu=0.01$.}
\end{subfigure}
\caption{\label{fig:PVplaneJ8} An illustration of the truncation process. 
For the chosen value of charges and $\Delta$ and $\delta\mu$, the 
extremal limit template is plotted and the cycle superimposed. If it intersects 
the disallowed region, the grid point is removed from consideration. Here, 
in (a), the black hole without any deficit is allowed, whereas the grid
point illustrated in (b) with acceleration and a deficit would be excluded.
}
\end{figure}

Given this insight, we investigated the effect of acceleration on rotating and
electrically charged black holes. We used both the method described in 
\S \ref{Efficiency}, and cross-checked against a discretised integration
of the heat flow around the cycle -- the method used in \cite{Hennigar:2017apu}.
First, the $(\Delta,\delta\mu)$ parameter space was discretised in a grid, and
the cycle checked against a template for the chosen values of $Q$ and $J$,
with the grid value of $\Delta$ and $\delta\mu$ to ensure that the cycle
remains within the allowed region of parameter space with non-negative
temperature (increasing $\Delta/C$ lowers the values of $Q$ and $J$ at which 
extremality occurs). Figure \ref{fig:PVplaneJ8} illustrates this template procedure.
Typically, charged cycles placed further from the origin in the $V$--$P$ plane can 
retain positive temperature over a larger range of $\Delta$, although the qualitative 
behaviour of efficiency remains the same.

\begin{figure}
\centering
\begin{subfigure}[b]{0.45\textwidth}
\includegraphics[width=\textwidth]{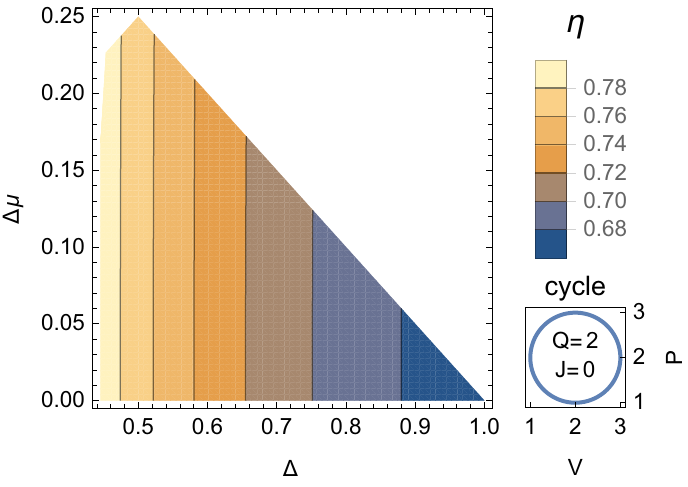}
\caption{\label{fig:PV2J0} Charged black hole, no rotation.}
\end{subfigure}
\quad
\begin{subfigure}[b]{0.45\textwidth}
\includegraphics[width=\textwidth]{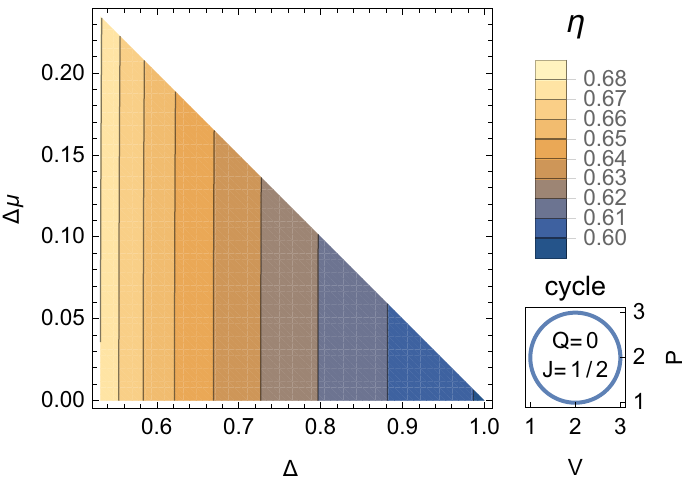}
\caption{\label{fig:PV2Q0} Rotating black hole, no charge.}
\end{subfigure}
\caption{\label{fig:PV2} Contour plots of the efficiency dependence in the
$(\Delta,\delta\mu)$ plane for a cycle centred on $P_0=V_0=2$ shown
for a charged black hole in (a) and a rotating black hole in (b). Due
to the placing of the cycle, and the chosen values of $J$ and $Q$, 
black holes on the cycle hit extremality for relatively large values of $\Delta$.
}
\end{figure}

\begin{figure}
\centering
\begin{subfigure}[b]{0.45\textwidth}
\includegraphics[width=\textwidth]{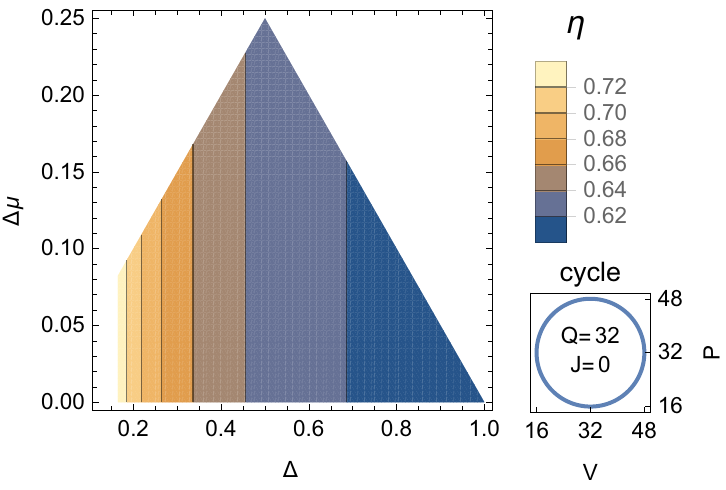}
\caption{\label{fig:PV32J0} Charged black hole, no rotation.}
\end{subfigure}
\quad
\begin{subfigure}[b]{0.45\textwidth}
\includegraphics[width=\textwidth]{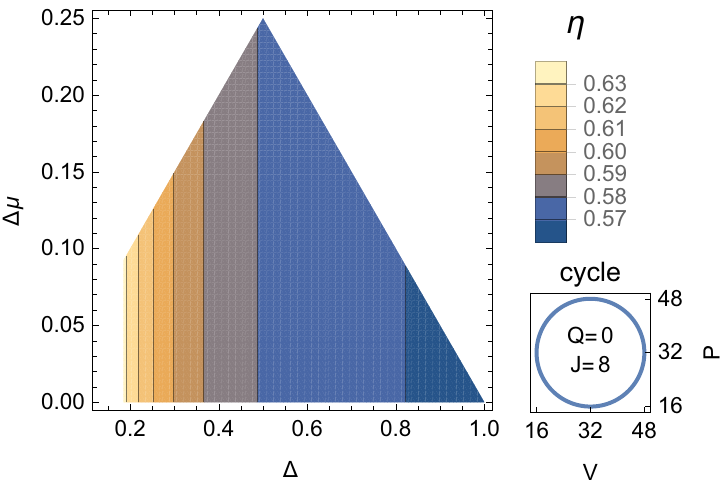}
\caption{\label{fig:PV32Q0} Rotating black hole, no charge.}
\end{subfigure} \\
\vskip 1cm
\begin{subfigure}[b]{0.45\textwidth}
\includegraphics[width=\textwidth]{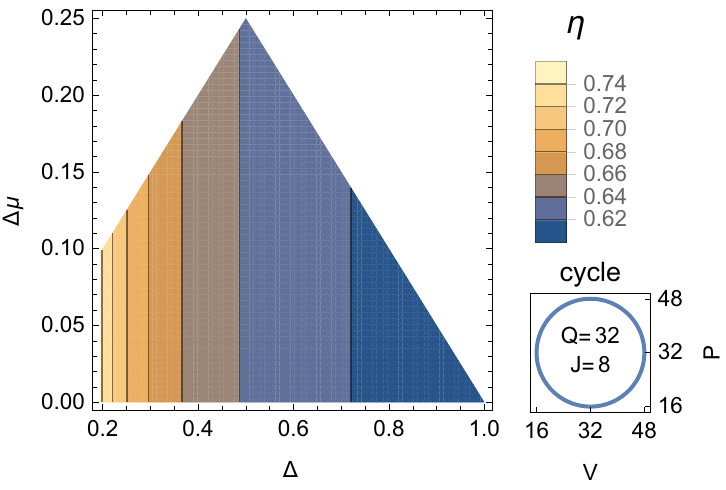}
\caption{\label{fig:PV32QJ} Charged and rotating black hole.}
\end{subfigure}
\begin{subfigure}[b]{0.45\textwidth}
\includegraphics[width=\textwidth]{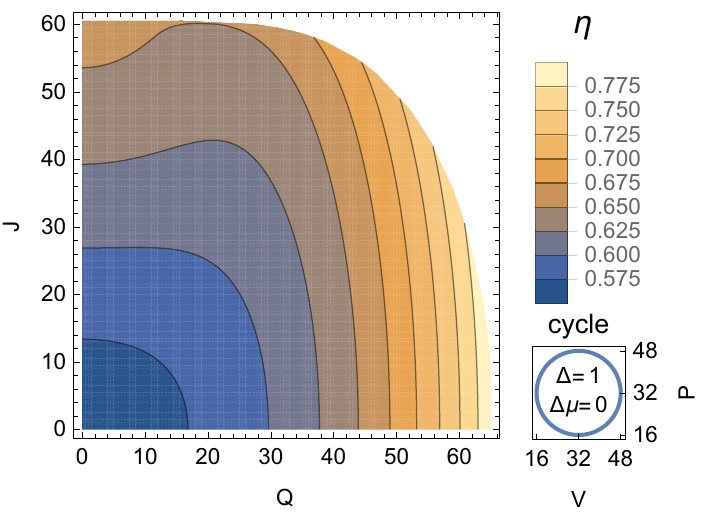}
\caption{\label{fig:PV32QJ0} The $Q$ and $J$ plane.}
\end{subfigure}
\quad
\caption{\label{fig:PV32} Contour plots of the efficiency dependence in the
$(\Delta,\delta\mu)$ plane for a cycle centred on $P_0=V_0=32$ shown
for black hole with either charge in (a), rotation in (b), or both in (c). 
In this case, the placing of the cycle allows a larger probe of the 
$(\Delta,\delta\mu)$ plane than figure \ref{fig:PV2}.
Subfigure (d) shows the dependence of efficiency on charge and rotation
for the pure black hole for reference. The truncation of parameter space
due to extremality is very clear.
}
\end{figure}

Next, the efficiency was calculated at each point on the grid in the allowed region,
and figures \ref{fig:PV2} and \ref{fig:PV32} illustrate the dependence
of efficiency on the deficit and acceleration for sample values of $Q,J$
in benchmarking cycles closer to, and further from,
the origin of the $(P,V)$ plane respectively. 
It is clear from the figures that the general conclusion is that adding charge
and rotation does not change the qualitative behaviour: the average conical 
deficit is dominant and acts to increase efficiency. If the geometry is restricted
(as in \cite{Zhang:2018vqs,Zhang:2018hms}) by imposing regularity on 
one axis, then as acceleration increases $\Delta$ will drop since the mean
deficit is increasing, and the efficiency will increase. It is misleading to assign
this effect to acceleration however, as figure \ref{fig:PV2} and \ref{fig:PV32} 
clearly show that significant changes in cycle efficiency are 
a consequence of the existence of a deficit, rather than the presence of 
acceleration.

\section{Discussion}
\label{concl}

We have explored the impact of conical deficits on the efficiency of
black hole heat engines. The result is that conical deficits give a marked 
improvement on the efficiency of black hole heat engines, whereas acceleration
has a relatively small effect, and tends to decrease efficiency. 
In the literature, when considering an accelerating
black hole, it is common to drive the acceleration by having a single cosmic
string segment emerging from one pole of the black hole. By restricting the tension
on the other pole to vanish, this leads to a constrained system - increasing 
acceleration necessarily increases the mean deficit of the spacetime, and it is this
mean deficit that has the largest impact on thermodynamics.

\begin{figure}
\centering
\includegraphics[width=0.5\textwidth]{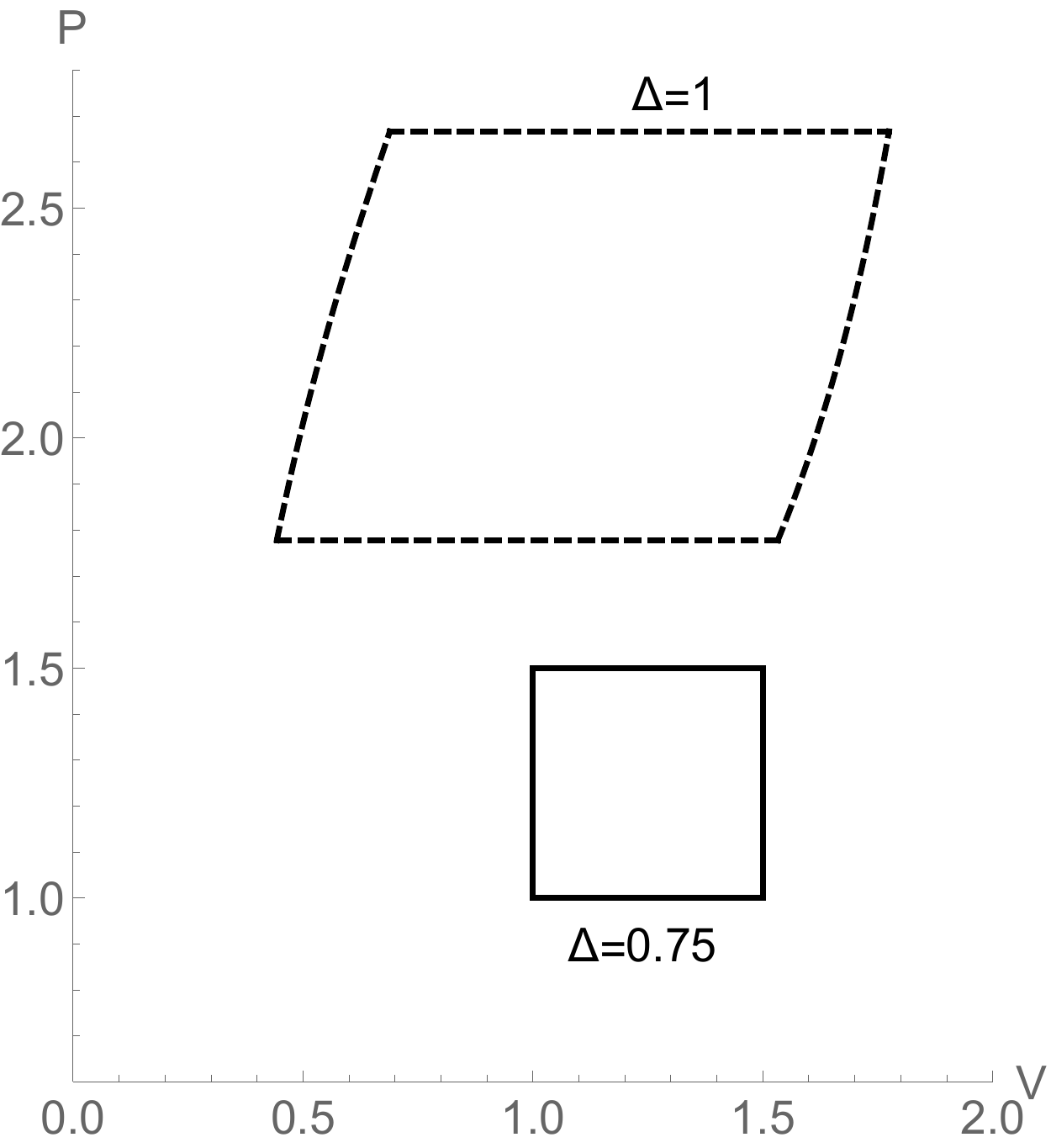}
\caption{An illustration of the impact of a conical deficit
on a square cycle of a rotating uncharged black hole ($J=1$). 
}
\label{fig:distortions}
\end{figure}

A simple understanding of why conical deficits increase efficiency can
be found by looking at the scaling \eqref{spscale}. This shows how adding a deficit
can be interpreted as a rescaling in the $(P,S)$ plane. The heat engines
are cycles in the $(P,V)$ plane, and while in inverting $S(V,P,...)$ the
other charges of the black hole come into play, there is no rescaling of these
charges ($Q$, $J$) so that one can still map a cycle with $\Delta$ into
a geometrically different cycle for the same black hole on the $(P,V)$ plane.
Figure \ref{fig:distortions} gives a simple illustration of how the deficit
can be thought of as a mapping between cycles on the $(P,V)$ plane for
a rotating black hole (chosen because it has $C_V\neq0$).
The small square cycle near the origin is mapped to a larger 
distorted almost-parallelogram further from the origin. The efficiency 
of the square cycle for the $J=1$ black hole with $\Delta=0.75$ is
mapped to the efficiency of the dashed cycle with $\Delta=1$.
Given that larger black holes are closer to the ideal gas limit (and
larger area cycles are typically more efficient), one can see the dual
drivers towards greater efficiency in this mapping.

A study of the efficiencies also reveals that the
placing of the benchmarking cycle can impact on the details of how the
efficiency varies. Typically, cycles placed closer to the origin have more variation
with acceleration. This is understood from a study of the `vanilla' C-metric
as coming from the order of magnitude of the $C-$term. It is also interesting to
note that rectangular benchmarking cycles bias against rotating black holes,
as efficiency largely decreases with rotation, whereas circular cycles show
a uniform increase. The efficiencies for each cycle are
\be
\eta_{\text{circ}} = \frac{\pi R^2}{\Delta M - V_0 \Delta P + {\cal S}/2 + C_1}
\;\;,\qquad
\eta_{\text{rect}}  = \frac{A}{\Delta M - V_L \Delta P }
\ee
where $\Delta M$ is the mass difference across the turning points of the cycle,
that typically increases with $J$ at fixed points in the $(P,V)$ plane.
$\Delta P$ likewise is the pressure difference, however, this has a very different
meaning for the rectangular and circular cycles. For the rectangular cycle,
the turning points are fixed, so the only variation is in $\Delta M$ in the
denominator that increases as $J$ increases, thus the efficiency drops. 
For the circular cycle however, the turning points shift as rotation is increased, 
so, not only is the contribution from the geometric terms on the denominator
lowered, but also the masses at the turning points, and $\Delta M$,
(although this is less obvious to see) hence efficiency increases.
This was precisely the motivation of Chakraborty and Johnson to
consider more general cycles in the $(P,V)$ plane \cite{Chakraborty:2016ssb,
Chakraborty:2017weq}.

\begin{wrapfigure}[13]{r}{0.5\textwidth}
\centering
\vskip -5mm
\includegraphics[width=0.5\textwidth]{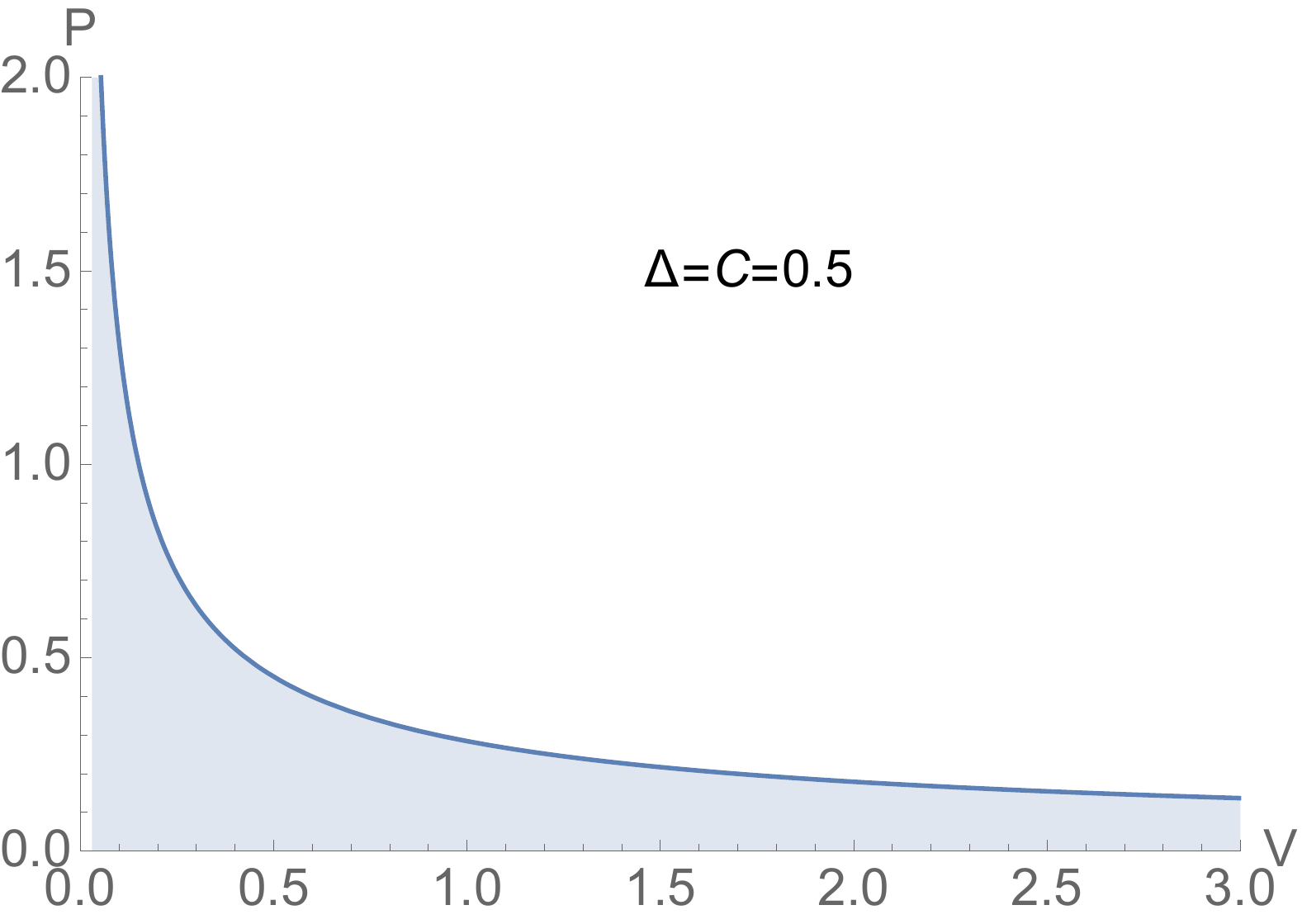}
\caption{
The exclusion region for a vanilla accelerating
black hole.}
\label{fig:exclude}
\end{wrapfigure}
Finally, although we did not explore extremely close to the origin of the 
$(P,V)$ plane, there are additional ``no-go'' regions once acceleration is 
introduced (see fig.\ \ref{fig:exclude}).
For small nonrotating black holes, it is possible for the enthalpy to vanish
due to the exothermic effect of acceleration in \eqref{ChRu}. This then 
means that the thermodynamic volume has a minimum, leading to an exclusion
region very near the origin. This is related to the ``snapping swallowtail''
phenomenon noted in 
\cite{Abbasvandi:2018vsh,Gregory:2019dtq,Abbasvandi:2019vfz}.
Thus, even though acceleration has little impact on the black hole
heat engine efficiency, it introduces some new subtleties for the
phase plane of the holographic heat engine.

\acknowledgments

We would like to thank David Kubiz\v n\'ak, Rob Mann and Fiona McCarthy for 
helpful discussions, and in particular Fiona McCarthy for making available
her mathematica notebook from \cite{Hennigar:2017apu}
that greatly facilitated our progress.
WA, H-ZC and EG are supported by the PSI program.
RG is supported in part by the STFC [consolidated grant ST/P000371/1], 
and in part by the Perimeter Institute. AS is supported by an STFC 
studentship, and would also like to thank Perimeter Institute for 
hospitality while this research was undertaken. 
Research at Perimeter Institute is supported by the Government of Canada 
through the Department of Innovation, Science and Economic Development 
and by the Province of Ontario through the Ministry of Research and Innovation.
Finally, we thank Camp Kintail for their hospitality and endless supply of
sustenance during the initiation of this project.

\end{document}